\documentclass{revtex4}

\usepackage{latexsym,natbib,amsmath}

\def\beq{\begin{equation}}
\def\eeq{\end{equation}}
\def\bea{\begin{eqnarray}}
\def\eea{\end{eqnarray}}

\def\d{{\partial}}

\def\dzeroh{{\hat\partial_0}}
\def\grad{{\nabla}}
\def\bgrad{{\bar \nabla}}

\newcommand{\christoffel}{\genfrac{\lbrace}{\rbrace}{0pt}{}{i}{j \; k}}
\newcommand{\Lie}{\pounds}
\newcommand{\M}{\mathcal{M}}

\begin{document}

\title{The Initial Value Problem Using Metric and Extrinsic Curvature}

\author{James W. York, Jr.}
\affiliation{Department of Physics, Cornell University, Ithaca, New York 14853}

\begin{abstract}
The initial value problem is introduced after a thorough review of the essential geometry.  The initial value equations are put into elliptic form using both conformal transformations and a treatment of the extrinsic curvature introduced recently.  This use of the metric and the extrinsic curvature is manifestly equivalent to the author's conformal thin sandwich formulation.  Therefore, the reformulation of the constraints as an elliptic system by use of conformal techniques is complete.
\end{abstract}

\maketitle

\section{Introduction}
Einstein's theory of gravity permits the use of arbitrary non-singular spacetime coordinates with a spacetime metric which is not specified beforehand, but is the desired solution.  This is what Einstein meant by his theory being ``generally covariant.''  It is a kind of gauge theory, where coordinate freedom plays the role, for example, of gauge transformation freedom on the four-potential $A_{\mu}$ in Maxwell's electrodynamics.  In Minkowski flat spacetime coordinates,
$A_{\mu} \rightarrow \bar A_{\mu} = A_{\mu} + \d_{\mu} f$, $f(t, \vec x)$ is the gauge function, and $\mu = 0, i$ where $i = 1,2,3$.

Gauge freedom comes with a price, a corresponding restriction on the field values on a constant-time hypersurface.  A detailed discussion is found, for example, in \cite{Tr}.  These restrictions constitute what is called the initial value problem for the theory in question.  In Maxwell's theory in the absence of sources, the constraint corresponding to gauge freedom is (with $c = 1$)
\[
\d E^i / \d x^i = 0 , \qquad E^i = - \d \phi / \d x^i - \d A^i / \d t
\]
where $A_\mu = (A_0, A_i) = (-\phi, A_i)$.  In acceptable physical theories, if the initial data constraints are satisfied at some time, then the remaining ``evolution equations'' carry the initial data forward (or backward) in time while preserving the constraints.  Thus one says that the ``Cauchy problem'' of the theory is well posed\index{well posed}.  That is, (1) the initial data constraints are satisfied and (2) the evolution equations propagate the initial data in such a way that the constraint continues to hold on every other spacelike hypersurface.  But a further problem also arises from gauge freedom.

Theories with gauge freedom, such as electromagnetism and general relativity, are said to be both ``overdetermined'' \index{overdetermined} and ``underdetermined.'' \index{underdetermined}  They are overdetermined because there are constraints at each time that limit the freedom of the variables that are propagated, the dynamical variables.  They are underdetermined because the gauge freedom means the equations of the theory cannot determine a fully unique solution.  By gauge transformations, some of the variables can be changed.  These changes do not alter the intrinsic physical meaning of a solution but they nevertheless can be vital in the \emph{description} and recognition of the solution.  That the problem of being overdetermined need be resolved at one time only (in principle), and that the gauge freedom in changing certain variables does not disturb either the feature just mentioned or the physical uniqueness of the problem are part and parcel of the well-posedness of a Cauchy problem.

In this article, the problem of setting up the constraints as an elliptic system using the inital data is treated.  (I will make no suggestion here about how to fix gauge variables.  But there \emph{is} something to say about just which variables are actually ``gauge'' variables.  One finds not the lapse and shift, but the ``densitized lapse'' \cite{Ash,AnYo} and the shift.)  One point of view is the initial value problem in the strictest sense, in which the objective is to construct directly the metric $g_{ij}$ of a spacelike hypersurface, or ``time slice'' $\M$, and the second fundamental tensor, or ``extrinsic curvature,'' $K_{ij}$ of the slice \cite{PfYo}.  The spatial tensor $K_{ij}$ does not depend on the behavior of the time coordinate ``$t$'' away from the slice.  That is, it is independent of any variables that would be needed to describe an infinitesimally nearby time slice $t = t_0 + \delta t$, if the initial slice is $t = t_0$.

The above caveat reminds us that there is also a ``thin-sandwich'' viewpoint in solving the initial value equations \cite{Yo1}.  Here, one introduces two infinitesimally close time slices: $t = t_0$ and $t = t_0 + \delta t$.  Roughly speaking, one says that the data given involve the spatial metric and its time derivative.  From there, one constructs indirectly the tensors $g_{ij}$ and $K_{ij}$ above.  Both viewpoints are useful and are equivalent, in the final analysis.

The equivalence of these two approaches using conformal methods, though clearly necessary for a full understanding of the initial-value part of the Cauchy problem, was established only relatively recently by this author (in 2001) and published in detail in \cite{PfYo}.  It also implies an outstanding result for stationary (rotating) black holes.  More generally, this result holds for \emph{any} slice in \emph{any} stationary spacetime \cite{Pf,PfYo}.  The result is that no ``transverse-traceless'' parts of $K_{ij}$ are needed to describe the initial data for a stationary spacetime .  This is interpreted to mean that there are no dynamical excitations present in a stationary spacetime, a result one would expect.  This kind of outcome has long been sought, first in using a flat background \cite{ADM}, and then by spatially covariant methods, first by Deser \cite{Des} and later in \cite{Yo2}.  The result was finally obtained only in \cite{PfYo}.

\section{Review of Some Basic Geometry}
The spacetime metric $g_{\mu\nu}$ will be written in the form
\beq
ds^2 = -N^2 (dt)^2 + g_{i j}(dx^i + \beta^i dt)(dx^j + \beta^j dt)
\eeq
where the spatial scalar $N$ is called the lapse function and $\beta^i$ is the (spatial) shift vector.  In a natural (coordinate) basis, $\beta_i = g_{0 i}$ ($\beta_i = g_{i j} \beta^j$ and $g_{i j}$, $g^{k l}$ are taken as the 3x3 inverses of one another; they are riemannian).  From this one can see that $\beta^i$ is a spatial vector and $\beta_i$ is a spatial one-form with respect to arbitrary spatial coordinate transformations provided the transformations are \emph{not} time-dependent.

The spacetime cobasis \index{cobasis} is
\beq
\theta^0 = dt \, , \qquad \theta^i = dx^i + \beta^i dt
\eeq
and the dual vector basis is
\beq
e_0 \equiv \d_0 = \d / \d t - \beta^i \d / \d x^i \, , \qquad e_i \equiv\d_i = \d / \d x^i .
\eeq
We see that $\d_0$ is a Pfaffian derivative \index{Pfaffian derivative} while $\d_t$ and $\d_i$ are natural derivatives.  The basis vector $\d_0$ can be generalized to the operator on spatial tensors
\beq
\dzeroh = \d_t - \Lie_\beta  \label{Eq:6}
\eeq
which, it should be noted, \emph{commutes with} $\d_i$ and propagates orthogonally to $t = const.$ slices.  It is obvious that $\d_i = \d / \d x^i$ and $\d_t = \d / \d t$ commute, because they are both natural derivatives.  What does not seem to be very well known at present, but in fact has been known for 50 or more years, is that the spatial Lie derivative \index{Lie derivative} $\Lie_\beta$, for any vector field $\beta^j$, \emph{commutes} with $\d / \d x^i$ when they act on tensors \emph{and} more general objects such as connections.  This result holds in an even more general form than the one we have just stated.  For an \emph{explicit} statement and discussion, see, for example, Schouten's \emph{Ricci Calculus} \cite{Sch}, p. 105, Eq. (10.17).  It is noteworthy that $\dzeroh$ acts orthogonally to $t = constant$ slices and that it is actually the \emph{only} time derivative that ever occurs in the 3+1 formulation of general relativity.

The connection coefficients \index{connection coefficients} in our ``Cauchy-adapted'' frame are given by
\beq
\omega^\alpha\mathstrut_{\beta\gamma} \Gamma^\alpha\mathstrut_{\beta\gamma} +
g^{\alpha\delta} C^{\epsilon}_{\delta ( \beta} g_{\gamma ) \epsilon} +
\frac{1}{2} C^\alpha\mathstrut_{\beta\gamma} \label{eq5}
\eeq
where $\Gamma$ denotes an ordinary Christoffel symbol, parentheses around indices denote the symmetric part (so $A_{(\beta\gamma)} = \frac{1}{2} (A_{\beta\gamma} + A_{\gamma\beta})$), and $C$ denotes the commutator \index{commutator}
\beq
[e_\beta, e_\gamma] = C^{\alpha}\mathstrut_{\beta\gamma} e_\alpha .
\eeq
Our spacetime covariant derivative \index{covariant derivative} convention associated with (\ref{eq5}) is
\beq
D_\alpha V^\gamma \d_\alpha V^\gamma + \omega^\gamma\mathstrut_{\alpha\beta} V^\beta .
\eeq
The only non-vanishing $C$'s are
\beq
C^i\mathstrut_{0 j} = - C^i\mathstrut_{j 0} = \d_j \beta^i .
\eeq

For the spatial covariant derivative we write
\beq
\grad_i V^j \d_i V^j + \gamma^j\mathstrut_{ik} V^k \d_i V^j +\Gamma^j\mathstrut_{ik} V^k .
\eeq
Because the shift $\beta^i$ is in the basis, the spacetime metric in the Cauchy basis that we use has no time - space components. A convenient consequence is that there is no ambiguity in writing $\Gamma^i_{j k}$.  For the spacetime metric determinant we will write $\det g_{\mu\nu} = -N^2 g$, $g = \det g_{ij}$; and $g_{ij}$ is here and hereafter considered as a 3x3 symmetric tensor.  For the usual $(- \det g_{\mu\nu})^{1/2}$ we will write $N \sqrt{g}$.
       
The $\omega$'s are given next.  Note that [MTW] write $\omega^\alpha\mathstrut_{\gamma\beta}$ where we have $\omega^\alpha\mathstrut_{\beta\gamma}$ for the same object, as in \cite{CbDw}.  Of course, this convention does not matter in a coordinate basis, where all of the connection coefficients are symmetric.
\bea
\omega^i\mathstrut_{jk} = \Gamma^i\mathstrut_{jk} (g_{\mu\nu}) &=& \Gamma^i\mathstrut_{jk} (g_{mn}) \\
\omega^i\mathstrut_{0j} = - N K^i\mathstrut_j +\d_j \beta^i , \quad
\omega^i\mathstrut_{j0} &=& - N K^i\mathstrut_j , \quad
\omega^0\mathstrut_{ij} = - N^{-1} K_{ij} \\
\omega^i\mathstrut_{00} = N \grad^i N , \quad
\omega^0\mathstrut_{0i} = \omega^0\mathstrut_{i0} &=& \grad_i \log{N} , \quad
\omega^0\mathstrut_{00} = \d_0 \log{N}
\eea
       
The Riemann tensor \index{Riemann tensor} satisfies the commutator
\beq
(D_\alpha D_\beta - D_\beta D_\alpha) V^\gamma (Riem)_{\alpha\beta}\mathstrut^\gamma\mathstrut_\delta V^\delta \label{eq10}
\eeq
where $(Riem)_{\alpha\beta}\mathstrut^\gamma\mathstrut_\delta$ would be denoted $(Riem)^\gamma\mathstrut_{\delta\alpha\beta}$ in [MTW].  Again, we are using the convention of \cite{CbDw}.
 
There are a number of possible definitions for the second fundamental tensor or extrinsic curvature tensor \index{extrinsic curvature tensor} $K_{ij}$. This does not measure curvature in the sense of Gauss or Riemann, where we can think of curvature as having dimensions $(length)^{-2}$.  The extrinsic curvature is a measure, at a point on a spatial slice, of the curvature of a spacetime geodesic relative to a spatial geodesic to which it is tangent at the point.  The dimensions are therefore $(length)^{-1}$. (See, for example, the Appendix of \cite{PfYo} for a detailed discussion.)  This is the same dimension as a connection symbol, and, in fact,
\beq
K_{ij} = - N \omega^0\mathstrut_{ij} \label{eq11}
\eeq
where $K_{kj} = K_{jk}$ and, from working out (\ref{eq11}), one finds for the metric velocity \index{metric velocity}
\beq
\dzeroh g_{ij} = - 2 N K_{ij} \d_t g_{ij} - (\grad_i \beta_j + \grad_j \beta_i) \label{eq12}
\eeq
where $\grad_i$ is the spatial covariant derivative with connection $\gamma^i\mathstrut_{jk} = \omega^i\mathstrut_{jk} = \Gamma^i\mathstrut_{jk}$; and the final term is, apart from sign, $\Lie_\beta \, g_{ij}$.

The Riemann tensor components are in accord with the Ricci identity (\ref{eq10})
\beq
(Riem)_{\alpha\beta}\mathstrut^\gamma\mathstrut_\delta \d_\alpha \omega^\gamma\mathstrut_{\beta\delta} -
\d_\beta \omega^\gamma\mathstrut_{\alpha\delta} +
\omega^\gamma\mathstrut_{\alpha\epsilon}
\omega^\epsilon\mathstrut_{\beta\delta} -
\omega^\gamma\mathstrut_{\beta\epsilon}
\omega^\epsilon\mathstrut_{\alpha\delta} -
C^\epsilon\mathstrut_{\alpha\beta}
\omega^\gamma\mathstrut_{\epsilon\delta} .
\eeq
The spatial Riemann tensor is denoted $R_{ij}\mathstrut^k\mathstrut_l$.  These curvatures are related by the Gauss - Codazzi - Mainardi equations \index{Gauss-Codazzi-Ricci equations} (see, for example, \cite{Sch})
\bea
(Riem)_{ijkl} &=& R_{ijkl} + (K_{ik} K_{jl} - K_{il} K_{jk}) \\
(Riem)_{0ijk} &=& N (\grad_j K_{ki} - \grad_k K_{ji}) \\
(Riem)_{0i0j} &=& N (\dzeroh K_{ij} + N K_{ik} K^k\mathstrut_j + \grad_i \d_j N)
\eea
One can likewise form and decompose the Ricci tensor, which has the definition
\beq
(Ric)_{\beta\delta} = (Riem)_{\alpha\beta}\mathstrut^\alpha\mathstrut_\delta .
\eeq
Thus,
\bea
(Ric)_{ij} &=& R_{ij} - N^{-1} \dzeroh K_{ij} + K K_{ij} -
2 K_{ik} K^k\mathstrut_j - N^{-1} \grad_i \d_j N \label{eq18} \\
(Ric)_{0j} &=& N (\d_j K - \grad_l K^l\mathstrut_j)
\equiv N \grad_l (\delta^l\mathstrut_j K - K^l\mathstrut_j) \label{eq19} \\
(Ric)_{00} &=& N (\d_0 K - N K_{ij} K^{ij} + \Delta N) \label{eq20}
\eea
where $\Delta N$ denotes the spatial ``rough'' or scalar Laplacian \index{scalar Laplacian} acting on the lapse function: $\Delta N \equiv g^{ij} \grad_i \grad_j N$.  The trace of $K_{ij}$ is $K$\index{trace of extrinsic curvature} and is called the ``mean curvature.''

It is important to know the spacetime scalar curvature, which we call $(TrRic)$:
\beq
(TrRic) = g^{\alpha\beta} (Ric)_{\alpha\beta} = g^{\alpha\beta} R_{\lambda\alpha}\mathstrut^\lambda\mathstrut_\beta \label{eq21}
\eeq
in the form $(N \sqrt{g} \equiv (- \det g_{\mu\nu})^{-1/2})$
\beq 
N \sqrt{g} (TrRic) = N \sqrt{g} (R + K_{ij} K^{ij} - K^2)
 - 2 \d_t (\sqrt{g} K) + 2 \d_i [\sqrt{g} (K \beta^i - \grad^i N)]
\eeq
where $R$ is the spatial scalar curvature, because the spacetime scalar curvature density is the Lagrangian density of the famous Hilbert action principle \cite{Hil}, explicitly modified in \cite{Yo5} to conform to the ADM action principle \cite{ADM}.  The scalar curvature itself will be needed later.  It is found from (\ref{eq21}) and (\ref{eq18}), (\ref{eq19}), and (\ref{eq20}) to be
\beq
(TrRic) = 2 N^{-1} \d_0 K - 2 N^{-1} \Delta N + (R + K_{ij} K^{ij} - K^2)
\eeq

\section{Einstein's Equations}
Einstein used his insights about the principle of equivalence and the principle of general covariance in arriving at the final form of his field equations.  As is now well known (if perhaps not explicitly documented?), his learning tensor analysis from Marcel Grossmann was an essential enabling step.  The equations, using the Einstein tensor \index{Einstein tensor}
\beq
(Ein)_{\mu\nu} \equiv G_{\mu\nu} \equiv (Ric)_{\mu\nu} - \frac{1}{2} g_{\mu\nu} (TrRic),
\eeq
are
\beq
G_{\mu\nu} = \kappa T_{\mu\nu}
\eeq
where $\kappa = 8 \pi G$, $c = 1$, $G$ = Newton's constant, and the stress-energy-momentum tensor of fields other than gravity (the ``source'' tensor) $T_{\mu\nu}$ must satisfy, as Einstein reasoned, in analogy to the conservation laws of special relativity,
\beq
\grad_\mu T^{\mu\nu} = 0
\eeq
corresponding to
\beq
\grad_\mu G^{\mu\nu} \equiv 0 ,
\eeq
which is an identity, the ``third Bianchi identity'' or the ``(twice) contracted Bianchi identity.'' \index{Bianchi identity}

Here we will consider only the vacuum theory.  This is non-trivial because the equations are non-linear (gravity acts as a source of itself) and because the global topology and/or possible boundary conditions are not prescribed by the equations.  The object of solving the equations is to find the metric.  In the 3+1 form of the equations, which is very close to a Hamiltonian framework, the object is to obtain $g_{ij}$ and $K_{ij}$, along with a useful specification of $N > 0$ and $\beta^i$ which, as we shall see shortly, are not determined by Einstein's equations.  For the vacuum case, one can use in four spacetime dimensions either of these two equations
\beq
G_{\mu\nu} = 0
\eeq
or
\beq
(Ric)_{\mu\nu} = 0 \label{eq28}
\eeq
because these equations are equivalent in three or more dimensions.  For completeness, we note that the form of equation (\ref{eq28}) with ``sources'' using the Ricci tensor is
\beq
(Ric)_{\mu\nu} = \kappa (T_{\mu\nu} - \frac{1}{2} T g_{\mu\nu}) \equiv \kappa \rho_{\mu\nu}
\eeq
where $T = T^\alpha_\alpha$, the trace of the stress-energy tensor.

\section{The 3+1-Form of Einstein's Equations}
A revealing way to write out the ten vacuum equations is to use both (Ric) and (Ein):
\beq
(Ric)_{ij} = 0, \qquad 2 N (Ric)^0_i = 0 , \qquad 2 G^0_0 = 0 \label{eq30}
\eeq
First, recall the geometric identity (\ref{eq12}).
\[
\dzeroh g_{ij} = - 2 N K_{ij}
\]
or
\beq
\d_t g_{ij} = - 2 N K_{ij}+ \Lie_\beta g_{ij}
= - 2 N K_{ij}+ (\grad_i \beta_j + \grad_j \beta_i) \label{eq31}
\eeq
From the first equation in (\ref{eq30}), one can obtain
\bea
\dzeroh K_{ij} &=& - \grad_i \d_j N + N(R_{ij} - 2 K_{il} K^l\mathstrut_j + K K_{ij}) \nonumber \\
&\equiv& \d_t K_{ij} + \Lie_\beta K_{ij} \nonumber \\
&\equiv& \d_t K_{ij} + \beta^l \grad_l K_{ij} + K_{il} \grad_j \beta^l + K_{lj} \grad_i \beta^l . \label{eq32}
\eea
The second and third equations in (\ref{eq30}) contain no terms $\d_t K_{ij}$ \index{extrinsic curvature time derivative} (\emph{i.e.}, no ``accelerations'' $\d_t \d_t g_{ij}$) and are, therefore, \emph{initial value constraints} \index{initial vaue constraints} on $g_{ij}$ and $K_{ij}$.  As previously mentioned, in this ``canonical''-like 3+1 form, there are no time derivatives of $N$ or of $\beta^i$.  In a second-order formalism, $\d_t N$ and $\d_t \beta^i$ would appear, as we see from (\ref{eq31}).  But no powers of $\dot N$ and $\dot \beta^i$ appear in (\ref{eq31}), the Lagrangian of Hilbert's action principle for the Einstein's equations.  We have thus an easy way of seeing that $\dot N$ and $\dot \beta^i$ are dynamically irrelevant.  We find from the second and third equations of (\ref{eq30}), respectively,
\bea
2 N R^0_i &\equiv& C_i = 2 \grad_j (K^j_i - K \delta^j_i) , \label{eq33} \\
2 G^0_0 &\equiv& C = K_{ij} K^{ij} - K^2 - R . \label{eq34}
\eea
These equations were derived in detail and displayed in \cite{Yo3}, \emph{without} a 3+1 splitting of the basis frames and coframes.  An arbitrary spacetime basis was used there in order to remove what some people regarded as the ``taint'' of using particular coordinates.  They are the standard 3+1 equations.  In regard to evolving $g_{ij}$ and $K_{ij}$, I do not claim that (\ref{eq31}) and (\ref{eq32}) are preferred for any other reason other than their maximum simplicity and absolute correctness when written in explicitly canonical form, using the variable $\pi^{ij}$, defined below, in place of $K_{ij}$.  I do not say anything here about the numerical properties of (\ref{eq31}) and (\ref{eq32}).

The canonical equations derived by Arnowitt, Deser, and Misner \cite{ADM} and by Dirac \cite{Di1,Di2}, are not equivalent to (\ref{eq32}) given above, even when written in the same formalism, that is, with $g_{ij}$ and $K_{ij}$.  This is because their equation of motion is $G_{ij} = 0$ rather than $R_{ij} = 0$.  Although $G_{\mu\nu} = 0$ and $R_{\mu\nu} = 0$ are equivalent, this is not true of spatial components.  Instead, one has the key identity \cite{AnYo}
\beq
G_{ij} + g_{ij} G^0_0 \equiv (Ric)_{ij} - g_{ij} g^{kl} (Ric)_{kl} ,
\eeq
or
\beq
G_{ij} + \frac{1}{2} g_{ij} C = (Ric)_{ij} - g_{ij} g^{kl} (Ric)_{kl} .
\eeq
Therefore, $G_{ij} = 0$ is not the correct equation of motion unless the constraint $C = 0$ also holds.

\emph{For the interested reader}, I remark that this means the Hamiltonian vector field of the ADM and Dirac canonical formalisms is not well-defined throughout the phase space.  There is an easy cure in the ADM approach, which is based on a canonical action principle.  When the metric $g_{ij}$ is varied, one must hold fixed the ``weighted'' or ``densitized'' lapse \index{densitized lapse} function $\alpha$=$g^{-1/2} N$, instead of just the scalar lapse $N$.  Thus, one carries out independent variations of $g_{ij}$, $\alpha$, $\beta^i$, and $\pi^{ij} = \kappa^{-1} g^{1/2} (K g^{ij} - K^{ij})$. \cite{Ash,AnYo}.

\section{Conformal Transformations}
A very useful technique for transforming the constraints $C_i$ and $C$ ((\ref{eq33}) and (\ref{eq34})) into a well posed problem involving elliptic partial differential equations is to use conformal transformations of the essential spatial objects $g_{ij}$, $K_{ij}$, $N$, and $\beta^i$.  Along the way we will again encounter the densitized lapse
\beq
\alpha \equiv g^{-1/2} N
\eeq
which has, as one sees, weight (-1): $N$ is a scalar with respect to spatial time-independent coordinate transformations (weight zero by definition).

The conformal factor \index{conformal factor} will be denoted by $\varphi$.  It will be assumed that $\varphi > 0$ throughout.  The conformal transformation \index{conformal transformation} is defined by its action on the metric
\beq
\bar g_{ij} = \varphi^4 g_{ij} . \label{eq35}
\eeq
It is called ``conformal'' because it preserves angles between vectors intersecting at a given point, whether one constructs the scalar product and vector magnitudes with $g_{ij}$ or $\bar g_{ij}$.  The power ``4'' in (\ref{eq35}) is convenient for three dimensions.  For dimension $n \geq 3$, the ``convenient power'' is $4 (n-2)^{-1}$ for the metric conformal deformation.  The neatness of this choice comes out most clearly in the relation of the scalar curvatures $\bar R = R(\bar g)$ and $R = R(g)$ below.
From (\ref{eq35}) and the fact that the spatial connection is simply the ``Christoffel symbol of the second kind $\christoffel$ '' which we denote by $\Gamma^i\mathstrut_{jk}$,
\beq
\Gamma^i\mathstrut_{jk} \equiv \christoffel = 
\frac{1}{2} g^{il} (\d_j g_{lk} + \d_k g_{lj} - \d_l g_{jk})
\eeq
we find that
\beq
\bar \Gamma^i\mathstrut_{jk} \Gamma^i\mathstrut_{jk} +
\frac{1}{2} \, \varphi^{-1} \,
(\delta^i_j \, \d_k \varphi + \delta^i_k \, \d_j \varphi
- g^{il} g_{jk} \, \d_l \varphi) .
\eeq
From $\bar \Gamma^i\mathstrut_{jk}$ we can find the relationship between $\bar R_{ij}$ and $R_{ij}$.
\beq
\bar R_{ij} = R_{ij} -
2 \varphi^{-1} \, \grad_i \d_j \varphi +
6 \varphi^{-2} \, (\d_i \varphi) (\d_j \varphi) -
g_{ij} [2 \varphi^{-1} \Delta \varphi +
2 \varphi^{-2} (\grad^k \varphi) (\d_k \varphi)]
\eeq
There is no need to derive the transformation of the Riemann tensor, for in three dimensions $R_{ijkl}$ can be expressed in terms of $g_{ij}$ and $R_{ij}$.  One can see that this must be so, for both the Ricci and Riemann tensors have six algebraically independent components.  The formula relating them has long been known (see below).  It is displayed for example in \cite{Yo3}.  This formula can be obtained from the identical vanishing of the Weyl conformal curvature tensor in three dimensions.  But riemannian three-spaces are \emph{not} conformally flat, in general.  The Weyl tensor in three dimensions is replaced by the Cotton tensor \index{Cotton tensor} \cite{Co}, which is conformally invariant and vanishes iff the three-space is conformally flat
\beq
C_{ijk} = \grad_j L_{ik} - \grad_k L_{ij}
\eeq
where
\beq
L_{ik} = R_{ik} - \frac{1}{4} R g_{ik} .
\eeq
Its dual \cite{Yo4}, using the inverse volume form $\epsilon^{mjk}$ on the skew pair [jk] and raising the index $i$ to $l$, in three dimensions, is a symmetric tensor $* \, C^{lm}$ with trace identically zero and covariant divergence identically zero.  The Cotton tensor has third derivatives of the metric and is therefore not a curvature tensor, but rather is a \emph{differential curvature tensor} with dimensions $(length)^{-3}$.  Therefore, the dual Cotton tensor \index{dual Cotton tensor} is
\beq
* \, C^{lm} = g^{li} \epsilon^{mjk} C_{ijk} .
\eeq
Under conformal transformations, we have that $\bar g^{ij} = \varphi^{-4} g^{ij}$, $\bar \epsilon_{ijk} = \varphi^{6} \epsilon_{ijk}$ (the volume three-form), $\bar C_{ijk} = C_{ijk}$, and $\bar \epsilon^{ijk} = \varphi^{-6} \epsilon^{ijk}$.  Therefore,
\beq
* \, \bar C^{ij} = \varphi^{-10} (* \, C^{ij}) . \label{eq45}
\eeq
The properties of $* \, C^{ij}$ hold for an entire conformal equivalence class, that is, for all strings of conformally related metrics as in (\ref{eq35}) for all $0 < \varphi < \infty$.  The divergence of $* \, C^{ij}$ is \emph{identically} zero, so it need not be surprising that
\beq
\bar \grad_j (* \, \bar C^{ij}) \varphi^{-10} \grad_j (* \, C^{ij}) \label{eq46}
\eeq
using the barred and unbarred connections to form the covariant divergence.  We see that (\ref{eq45}) is the natural conformal transformation law for symmetric, traceless type $\binom{2}{0}$ tensors in three dimensions, whose divergence may or may not vanish.  Note that in obtaining (\ref{eq46}), the scaling (\ref{eq45}) was used.  But \emph{no} properties of $* \, C^{ij}$ were employed except the symmetry type of the tensor representation: symmetric with zero trace.

We now pass to the famous formula for the conformal transformation of the scalar curvature \index{conformal transformation of the scalar curvature} $R$, which occurs in the constraint equation.
\beq
\bar R = \varphi^{-4} R - 8 \varphi^{-5} \Delta \varphi .
\eeq

So far every transformation has followed from the defining relation $\bar g_{ij} = \varphi^4 g_{ij}$.  A glance at both the constraints (\ref{eq33}) and (\ref{eq34}) shows that we must deal with $K_{ij}$.  The method here can be deduced by writing $K_{ij}$ as
\beq
K^{ij} = A^{ij} + \frac{1}{3} K g^{ij}
\eeq
where $A^{ij}$ is the traceless part of $K^{ij}$.  We treat $A^{ij}$ and $K$ differently because $A^{ij}$ and $K g^{ij}$ can be regarded as different irreducible types of symmetric two-index tensors.  The conformal method was introduced into this problem by Lichnerowicz \cite{Lich} who took $K = 0$.  But this is too restrictive, and even then the simplified ``momentum constraint'' (\ref{eq33}) was not solved.  A very useful result was given in \cite{Yo2} and it was used for many years to solve the momentum constraint.  A better method yet was displayed in \cite{PfYo} and is given below.

Suppose an overbar denotes a solution of the constraints and the corresponding object without an overbar denotes a ``trial function.''  The strategy is to ``deform'' the trial objects conformally into barred quantities, that is, into solutions.

\section{An Elliptic System}
We write (\ref{eq33}) and (\ref{eq34}) in the barred variables as
\bea
\bgrad_j \bar A^{ij} - \frac{2}{3} \bar g^{ij} \d_j \bar K = 0 \label{eq50} \\
\bar A_{ij} \bar A^{ij} - \frac{2}{3} \bar K - \bar R = 0 \label{eq51}
\eea
Conformal transformations for objects that are purely concomitants of $\bar g_{ij}$ (or $g_{ij}$) are derived as above in a straightforward manner.  But the extrinsic curvature variables have to be handled with a modicum of care.  The transformations obtained by extending $\bar g_{ij} = \varphi^4 g_{ij}$ to all of the spacetime metric variables is not appropriate because the view that spacetime structure is primary is not helpful in a situation, as here, where there is as yet \emph{no} spacetime.

We begin with $\bar K$.  We hold it fixed because its inverse in the simpler cosmological models is the ``Hubble time,'' \index{Hubble time} without a knowledge of which the epoch is not known.  Data astronomers obtain from different directions in the sky, or at different ``depths'' back in time are basically correlated and they fix $\bar K$ implicitly.  Therefore, I long ago adopted the rule \cite{Yo5} of fixing the ``mean curvature'' \index{fixing the mean curvature} $\bar K$
\beq
\bar K = K
\eeq
under conformal transformations.  Thus it is known \emph{a priori}.

What to do about the the symmetric tracefree tensor $\bar A^{ij}$?  The prior discussion of $* C^{ij}$ indicates the transformation
\beq
\bar A^{ij} = \varphi^{-10} A^{ij} \label{eq53}
\eeq
But symmetric tensors ``$\bar T^{ij}$'' have, in a curved space, three irreducible types that are formally $L^2$-orthogonal.  One is the trace $(\bar g^{ij} \bar T^k_k)$, another is like $* C^{ij}$, that is, a part with vanishing covariant divergence.  Finally, a symmetric tracefree tensor can be constructed from a vector
\beq
(\bar L X)^{ij} (\bgrad^i X^j + \bgrad^j X^i - \frac{2}{3} \bar g^{ij} \bgrad_l X^l) , \label{eq54}
\eeq
the ``conformal Killing form'' of $X^i$.  (I have not found other constructions that are sufficiently well-behaved under conformal transformations to be useful in this problem.)  It vanishes iff $X^i$ is a conformal killing vector of $\bar g_{ij}$.  Therefore, $X^i$ is a conformal killing vector of every metric conformal to $\bar g_{ij}$.  Therefore,
\beq
\bar X^i = X^i , \qquad \bar g_{ij} = \varphi^4 g_{ij} ,
\eeq
and
\beq
(\bar L X)^{ij} = \varphi^{-4} (L X)^{ij} ,
\eeq
which \emph{misses} obeying our ``rule'' (\ref{eq45}) or (\ref{eq53}).  For a long time, the mismatched powers required a work-around to obtain an \emph{ansatz} for solving the constraints \index{solving the constraints} \cite{Yo3}, but I arrived at a simple solution fairly recently (2001) \cite{Yo4, PfYo}.  The vectorial part (\ref{eq54}) needs a weight factor and a corresponding change in the measure of orthogonality.  The solution seems to me simple and beautiful.

Recall our statement that the densitized lapse $\alpha$ is the preferred undetermined multiplier (rather than the lapse $N$) in the action principle leading to 3+1 (or canonical) equations of motion.  See \cite{AnYo} where this is made perfectly clear.  This is not to say anything about the ``best'' form of the $\dzeroh K_{ij}$ (or $\dzeroh \pi^{ij}$) equation of motion for calculational purposes.  In fact, the system for $\dzeroh g_{ij}$ and $\dzeroh K_{ij}$ is only ``weakly hyperbolic.''  (For me, the presence of the spatial Ricci tensor in the equation for $\dzeroh K_{ij}$ is what I have always regarded as ``the problem.'')  But I do say that \emph{only} this form gives a Hamiltonian vector field well-defined in the entire momentum phase space.

To proceed, we note that one does not scale undetermined multipliers\index{undetermined multipliers}.  Therefore
\beq
\bar \beta^i = \beta^i , \qquad \bar \alpha = \alpha .
\eeq
But because $\bar \alpha = \bar g^{-1/2} \bar N$ and $\bar g^{1/2} = \varphi^6 g^{1/2}$, then $\bar N = \bar g^{1/2} \alpha = g^{1/2} \alpha$ implies
\beq
\bar N = \varphi^6 N . \label{eq58}
\eeq

I have known that the transformation (\ref{eq58}) was useful since 1971.  But, thinking that $N$ was an undetermined multiplier - a lowly ``C-number,'' independent of the dynamical variables, in Dirac's well-known parlance - I did not use (\ref{eq58}).  Then I learned about the densitized lapse and saw its role in the action principle.  It then dawned on me that (\ref{eq58}) was correct all along.  It made its first appearance in the conformal thin sandwich problem \cite{Yo1}.

The lapse becomes, thus, a dynamical variable \cite{Ash, AnYo, PfYo}.  A look at (\ref{eq58}) and at the relation between $\d_t \bar g_{ij}$ and $\bar K_{ij}$ gives us the scalar weight factor $(-2 \bar N)^{-1}$ in the decomposition of $\bar A^{ij}$
\beq
\bar A^{ij} = \bar A^{ij}_{(\delta)} +
(-2 \bar N)^{-1} (\bar L X)^{ij} . \label{eq59}
\eeq
The subscript $(\delta)$ indicates that the covariant divergence of $\bar A^{ij}_{(\delta)}$ vanishes.  Note that (\ref{eq59}) does \emph{not} mean that the extrinsic curvature is sensitive to $N$.  It is not.  What it \emph{does} mean is that the identification of the divergence-free and trace-free part of the extrinsic curvature is, in part, dependent on $N$.  Also note that the two parts of $\bar A^{ij}$ are formally $L^2$-orthogonal both before and after a conformal transformation, with the geometrical \emph{spacetime measure}\index{spacetime measure}
\beq
(- \det g_{\mu\nu})^{1/2} = N g^{1/2}
\eeq
instead of the spatial measure.  Therefore, we have
\bea
\int \bar A^{ij}_{(\delta)} [(-2 \bar N)^{-1} (\bar L X)^{kl}] \bar g_{ik}
\bar g_{jl} (\bar N \bar g^{1/2}) d^3x \nonumber \\
= \int A^{ij}_{(\delta)} [(-2 N)^{-1} (L X)^{kl}] g_{ik}
g_{jl} (N g^{1/2}) d^3x \label{eq61}
\eea
Upon integration by parts, with suitable boundary conditions, or no boundary, each of the integrals (\ref{eq61}) vanishes.

We construct $\bar A^{ij}_{(\delta)}$ or $\bar A^{ij}_{(\delta)}$ by extracting from a freely given symmetric tracefree tensor $\bar F^{ij} = \varphi^{-10} F^{ij}$  its transverse-tracefree part, which will be our $A^{ij}_{(\delta)}$
\beq
F^{ij} = A^{ij}_{(\delta)} + (-2 N)^{-1} (L Y)^{ij}
\eeq
with
\beq
\grad_j [(-2 N)^{-1} (L Y)^{ij}] = \grad_j F^{ij}
\eeq
The momentum constraints
\beq
\bgrad_j \bar A^{ij} - \frac{2}{3} \bar g^{ij} \d_j K = 0 \label{eq64}
\eeq
become, with $Z^i = X^i - Y^i$,
\beq
\grad_j [(-2 N)^{-1} (L Z)^{ij}] = \grad_j F^{ij} -
\frac{2}{3} \varphi^6 g^{ij} \d_j K .
\eeq
The solution for $Z^i$, with given $N$, will give the parts of $\bar K_{ij}$,
\bea
& &\bar A^{ij} = \varphi^{-10} [F^{ij} + (-2 N)^{-1} (L Z)^{ij}] , \\
\nonumber \\
& &\frac{1}{3} \bar K \bar g^{ij} = \frac{1}{3} \varphi^{-4} K g^{ij} .
\eea
However, (\ref{eq64}) contains $\varphi$ and is coupled to the ``Hamiltonian constraint'' (\ref{eq51}) unless the ``constant mean curvature'' (CMC) condition $K = constant$ (in space, $\d_j K = 0$) can be employed, as introduced in \cite{Yo5}.  This includes maximal slicing $K = 0$ \cite{Lich}.  (Lichnerowicz did not propose the CMC condition as claimed in \cite{TipMar}.)

Gathering the transformations for $\bar R$ and $\bar K^{ij}$ enables us to write the Hamiltonian constraint as the general relativity version of the Laplace-Poisson equation
\beq
\Delta \varphi \frac{1}{8} [R \varphi +
(F_{ij} + (-2 N)^{-1} (L Z)_{ij})^2 \varphi^{-7} -
\frac{2}{3} K^2 \varphi^5] = 0 .
\eeq

Suppose, for example, we choose $N = 1$.  Then
\beq
\bar N = \varphi^6 = \bar g^{1/2} (g^{-1/2}) .
\eeq
We are certainly entitled to have chosen $g_{ij}$ such that $g^{1/2} = 1$, without loss of generality.  Thus we recover Teitelboim's gauge for the lapse equation
\beq
\bar N = \bar g^{1/2}
\eeq
in his noted paper on the canonical path integral in general relativity \cite{Teit}.

A bit more generally, if we choose
\beq
\dzeroh g^{1/2} = 0 ,
\eeq
we see that $\bar N$ automatically satisfies the time gauge equation used by Choquet-Bruhat and Ruggeri \cite{CbRu}.

The constraints have the same form as they do in the thin sandwich formulation: see \cite{Yo1}.  Therefore, the space of solutions has the properties obtained in \cite{CbIsYo}.

Finally, if we write out from the formula for $\d_t \bar g_{ij}$ its tracefree part $\bar u_{ij}$, or the velocity of the conformal metric, we obtain
\beq
\bar u_{ij} = -2 \bar N \bar F_{ij} + [\bar L (\bar Z + \bar \beta)]_{ij}
\eeq
with $(\bar Z + \bar \beta)_j \equiv \bar g_{ij} (Z^i + \beta^i)$.  This has the form of the solution in the conformal thin sandwich problem.  The choice of shift $\beta^i = - Z^i$ is possible here and renders a simple final form.

\section*{Acknowledgments}
This work was supported by the National Science Foundation of the U.S.A., PHY-0311817.  The author thanks A. Lundgren for his invaluable assistance in preparing the manuscript.  For recent helpful conversations, he thanks Y. Choquet-Bruhat, J. Isenberg, and H. Pfeiffer.

\end{document}